\documentclass[superscriptaddress,aps,prd,twocolumn,nofootinbib]{revtex4}
\usepackage[dvips]{graphicx}
\usepackage{amsfonts}
\usepackage{slashed}
\usepackage{xcolor}
\usepackage{amsmath}
\usepackage{amssymb}
\usepackage{hyperref}
\newcommand{\be}{\begin{equation}}
	\newcommand{\en}{\end{equation}}
\newcommand{\ba}{\begin{eqnarray}}
	\newcommand{\ea}{\end{eqnarray}}
\newcommand{\bea}{\begin{eqnarray}}
	\newcommand{\eea}{\end{eqnarray}}

\newcommand{\sech}{{\rm sech}}

\begin{document}

\title{Braneworlds in bumblebee gravity}

\author{M. A. Marques\,
}
\email[]{marques@cbiotec.ufpb.br}
\affiliation{Departamento de Biotecnologia, Universidade Federal da Para\'{\i}ba,\\ 58051-900, Jo\~ao Pessoa, Para\'{\i}ba, Brazil}
			
\author{R. Menezes\,
}
\email[]{rmenezes@dcx.ufpb.br}
\affiliation{Departamento de Ci\^{e}ncias Exatas, Universidade Federal da Para\'{\i}ba,\\
			58927-000, Rio Tinto, Para\'{\i}ba, Brazil}
\author{A. Yu. Petrov\,
}
	\email[]{petrov@fisica.ufpb.br}
	\affiliation{Departamento de F\'{\i}sica, Universidade Federal da Para\'{\i}ba,\\
		Caixa Postal 5008, 58051-970, Jo\~ao Pessoa, Para\'{\i}ba, Brazil}
\author{P. J. Porf\'{\i}rio\,
}
\email[]{pporfirio@fisica.ufpb.br}
	\affiliation{Departamento de F\'{\i}sica, Universidade Federal da 
		Para\'{\i}ba,\\
		Caixa Postal 5008, 58051-970, Jo\~ao Pessoa, Para\'{\i}ba, Brazil}

	
\begin{abstract}
We investigate thick-brane solutions within the five-dimensional bumblebee gravity in the presence of a real scalar field. Specifically, we implement the Lorentz symmetry breaking scenario within this context and obtain brane-like structures. Since the contribution of the bumblebee field is expected to be weak, we solve the field equations in the small-parameter regime. In this situation, we develop a first-order framework to describe the brane. The results show that the function which drives the bumblebee field may engender a lumplike structure whose shape depends on the parameters. On one hand, the effect of the non-minimal coupling, controlled by  $\xi$, between the bumblebee field and gravity shifts the field from the vacuum expectation value. On the other hand, the aether parameter, $\beta$, is responsible for modifying the solution inside the brane.
\end{abstract}
	
	
	\maketitle

\section{Introduction}
\label{sec:intro}
 
One of the most interesting problems involving studies of Lorentz symmetry breaking is its consistent incorporation within the gravity context. First steps along this way were done in the seminal work \cite{KosGra} where a generalization of the Lorentz-Violating Standard Model Extension (LV SME), originally introduced in the particle physics sector of SM \cite{Colladay1, Colladay2}, for the gravitational sector was proposed. In particular, the bumblebee gravity model \cite{KosGra},  which is constructed with the use of a dynamical vector field responsible for the Lorentz symmetry breaking,  is the most convenient way for incorporating spontaneous Lorentz symmetry breaking in curved spaces. Indeed, this model is based on choosing a particular vacuum expectation value (VEV) of the dynamical bumblebee field, thus introducing a preferred frame in a natural manner. This mechanism is the most suitable one for introducing Lorentz violation in a curved space-time since the usual manner of explicit Lorentz symmetry breaking is known to generate severe problems (namely, arising of infinite towers of covariant derivatives of "constant" LV parameters like in \cite{Shapiro} or presence of so-called no-go constraints allowing to eliminate such towers but very difficult to satisfy \cite{KosLiGrav}). Therefore, the bumblebee gravity appears to be a very promising way to conciliate Lorentz symmetry breaking in a curved space-time. 

Regarding bumblebee gravity solutions, the consistency of various known metrics has been confirmed. The paradigmatic examples are the static spherically symmetric metric \cite{Bert, Poulis:2021nqh}, cosmological FRW metric \cite{Para}, black hole metric \cite{Casana, ourPP}, G\"{o}del \cite{godelbumb} and G\"{o}del-type metrics \cite{godelbumb1}. Therefore, it is natural to apply the bumblebee gravity within other physically relevant situations.

In the curved spacetime, the braneworld scenario has been investigated over time due to its many applications within the elementary particle and gravity context \cite{RS1,RS2,garriga}. In this direction, one may consider the inclusion of scalar fields in the action to modify the brane, which gets a thick profile when it is modelled by kinklike structures \cite{dewolfe,cvetic,csaki,gremm,brito,kobayashi,dzhunushaliev}. This approach supports a first order formalism that allows for branes with distinct profiles; see Refs.~\cite{b1,b2,b3,b4,b5,b6,b7,b8,b9}. In particular, in Ref.~\cite{b3}, models of asymmetric thick branes were investigated, showing that the asymmetry may arise due to the interpolation of different geometries, because of the asymptotic behavior of the scalar field solution or as a consequence of the asymmetric character of the scalar potential that gives rise to the brane. Also, in Ref.~\cite{b4}, it was shown that the compactification of the scalar field solution leads to the hybrid brane, which engenders a thick (thin) profile for points inside (outside) a compact space.

The study of Lorentz-violating theories in spacetimes with a single extra dimension has been performed in Refs.~\cite{CarrollTam,aether}, where mostly flat space-time or weak gravity were considered. Therefore, considering a curved five-dimensional space-time is rather natural in this context. 

In this manuscript, we investigate the presence of five-dimensional thick branes within the bumblebee gravity, seeking departures from the usual GR solution. In the section \ref{sec2}, we write down the action and equations of motion of the model. In the section \ref{sec3}, we develop a first-order formalism in order to obtain solutions for the field equations. Finally, in the section \ref{sec4}, we discuss the results.

\section{Classical action and equations of motion}\label{sec2}

Our starting point is the five-dimensional generalization for the action proposed in \cite{Casana}:
\begin{equation}\label{kh}
	\begin{aligned}
		S=\int d^5x \sqrt{|g|}\bigg(&\frac{1}{2\kappa}(R+\xi B^{\mu}B^{\nu}R_{\mu\nu})-\frac{1}{4}B_{\mu\nu}B^{\mu\nu} \\
		 &-V(B^{\mu}B_{\mu}\pm b^2)+{\cal L}_M (g_{\mu\nu},\psi, B_{\mu})\bigg),
	\end{aligned}
\end{equation}
where $g_{\mu\nu}$, $B_{\mu}$ and $\psi$ are the metric tensor, the bumblebee field and a set of matter fields, respectively. The parameter $\xi$ controls the non-minimal coupling of the bumblebee field with the gravity via the Ricci tensor and $\kappa=8\pi G^{(5)}$. Also, $B_{\mu\nu}=(dB)_{\mu\nu}$ represents the field strength and the potential $V(B^{\mu}B_{\mu}\pm b^2)$ is responsible for the spontaneous Lorentz symmetry breaking. Actually, a non-trivial minimum of the potential leads the bumblebee to develop a non-zero VEV, we say $<B_{\mu}>=b_{\mu}$. Furthermore, $b^2 =g^{\mu\nu}b_{\mu}b_{\nu}$ is the square norm of the VEV. The matter fields represented by $\psi$ are minimally coupled to gravity through ${\cal L}_M (g_{\mu\nu},\psi, B_{\mu})$.

Varying the aforementioned action with respect to the metric, we are able to find the gravitational field equations, namely, 
\bea
\label{eqgra}
R_{\mu\nu}-\frac{1}{2}Rg_{\mu\nu}=\kappa (T^B_{\mu\nu}+T^M_{\mu\nu}),
\eea
in which $T^M_{\mu\nu}$ is the energy-momentum tensor of the usual matter sources (scalar field, relativistic fluid, etc), and $T^B_{\mu\nu}$ is the energy-momentum tensor of the pure contributions of the bumblebee field. They, respectively, look like
\begin{align}
\label{tem}
T^{M}_{\mu\nu}&=-\frac{2}{\sqrt{|g|}}\frac{\delta(\sqrt{|g|}\mathcal{L}_{M})}{\delta g^{\mu\nu}},\\
T^B_{\mu\nu}&=-B^{\mu\alpha}B_{\alpha}^{\phantom{\alpha}\nu}-\frac{1}{4}g_{\mu\nu}B^{\lambda\rho}B_{\lambda\rho}-Vg_{\mu\nu}+
2V_X B_{\mu}B_{\nu}+\nonumber\\
&+\frac{\xi}{\kappa}\Big[
\frac{1}{2}B^{\alpha}B^{\beta}R_{\alpha\beta}g_{\mu\nu}-B_{\mu}B^{\alpha}R_{\alpha\nu}-B_{\nu}B^{\alpha}R_{\alpha\mu}+\nonumber\\
&+\frac{1}{2}\nabla_{\alpha}\nabla_{\mu}(B^{\alpha}B_{\nu})+\frac{1}{2}\nabla_{\alpha}\nabla_{\nu}(B^{\alpha}B_{\mu})-\frac{1}{2}\nabla^2(B_{\mu}B_{\nu})\nonumber\\
&-\frac{1}{2}g_{\mu\nu}\nabla_{\alpha}\nabla_{\beta}(B^{\alpha}B^{\beta})
\Big],
\end{align}
where $V_X=\frac{dV}{dX}$, with $X=B^{\mu}B_{\mu}\pm b^2$.
We note that the analogous equations within the metric-affine formalism were obtained in \cite{classMetAf,qMetAf1,qMetAf2}.

Now, varying the action (\ref{kh}) with respect to the bumblebee field to get their respective equations of motion, by doing so, one obtains
\bea
\label{eqmat}
\nabla^{\mu}B_{\mu\nu}=J^{M}_{\mu}-2V_X B_{\nu}-\frac{\xi}{\kappa}B^{\mu}R_{\mu\nu},
\eea
with $J^{M}_{\mu}=\delta\mathcal{L}_{M}/\delta B^{\mu}$ denoting the matter current which is related to the coupling between the matter sources and the bumblebee field.

So, our goal consists in investigating how the presence of the bumblebee field modifies the braneworld scenario \cite{RS1,RS2,dewolfe} with line element given by
\bea
ds^2=e^{2A(y)}\eta_{ab}dx^adx^b-dy^2,
\label{brane}
\eea
where Latin indices label ``parallel''  coordinates to the brane ($a,b=0,1,2,3$), $y$ is the coordinate of the extra dimension  and $A(y)$ is the warp function. To do so, one must solve equations \eqref{eqgra} and \eqref{eqmat} in the background (\ref{brane}). The first step is to take an ansatz for the bumblebee field and potential. In particular, we consider $B^{\mu}=[0,0,0,0, B(y)]$, i.e., its only non-trivial component is along the extra dimension and also only $y$-dependent. As for the potential, one can choose the usual Mexican hat one, $V(B^{\mu}B_{\mu}\pm b^2)=\frac{\lambda_B}{4}(B^{\mu}B_{\mu}\pm b^2)^2$, where $\lambda_B$ is a positive parameter and $b$ is constant. Moreover, we consider a scalar field, $\phi$, with an exclusive dependency on the $y$ coordinate, as the only matter content. Then,
\begin{equation}\label{sm}
    \mathcal{L}_{M}=-\left(\frac{1}{2}\partial_{\mu}\phi\partial^{\mu}\phi-U(\phi)+\beta B^{\mu}B^{\nu}\partial_{\mu}\phi\partial_{\nu}\phi\right),
\end{equation}
in which $U(\phi)$ is the potential of the scalar field and $\beta$ is the coupling constant controlling the non-minimal interaction between the bumblebee and scalar fields. In this case, the current below Eq.~\eqref{eqmat} becomes $J^M_\mu=2\beta B^\nu \partial_\mu\phi\partial_\nu\phi$. The interaction term in the above expression is the aether term for the scalar field proposed in \cite{CarrollTam,aether}, with the bumblebee field $B^{\mu}$ playing the role of the LV vector. Such a scalar field satisfies the following equation of motion
\begin{equation}\label{eomphig}
	\Box\phi+U_\phi+2\beta \nabla_{\mu}\left(B^{\mu}B^{\nu}\partial_\nu\phi\right)=0,
\end{equation}
with $U_\phi=dU/d\phi$. Note that Eq. (\ref{eomphig}) is the Klein-Gordon equation added up to a potential and a sourcing term, which stems from the aether term in the action (\ref{sm}).  The energy-momentum tensor of the scalar field is
\begin{equation}
\begin{aligned}
	T^{(\phi)}_{\mu\nu}&=\partial_{\mu}\phi\partial_{\nu}\phi-\frac{1}{2}g_{\mu\nu}\partial_{\alpha}\phi\partial^{\alpha}\phi+U(\phi)g_{\mu\nu}\\
	&+\beta\left[4B^{\lambda}B_{(\mu}\partial_{\nu)}\phi\partial_{\lambda}\phi - g_{\mu\nu}B^{\alpha}B^{\lambda}\partial_{\alpha}\phi\partial_{\lambda}\phi\right].
\end{aligned}
\end{equation}
In this scenario, the non-null components of Eq.~\eqref{eqgra} for the line element in Eq.~\eqref{brane} are
\begin{subequations}\label{einsteineq}
	\begin{equation}
\begin{aligned}
	&6{A^\prime}^2-\frac{\kappa}{2}\,{\phi^\prime}^2 +\kappa U - \frac{\kappa\lambda_B b^4}{4} - \frac{\kappa\lambda_B b^2}{2} B^2 +\frac{3\kappa\lambda_B}{4} B^4 \\
	&-2\kappa\xi B^2{A^\prime}^2-4\kappa\xi BB^\prime A^\prime + 4\kappa\xi B^2 A^{\prime\prime} + 6\kappa\beta B^2 {\phi^\prime}^2=0,
	\end{aligned}
\end{equation}
\begin{equation}
	\begin{aligned}
		&3A^{\prime\prime} + \kappa {\phi^\prime}^2 + \kappa \lambda_B b^2 B^2 -\kappa\lambda_B B^4 -4\kappa\beta B^2{\phi^\prime}^2- \kappa\xi {B^\prime}^2\\
		&-4\kappa\xi B^2 {A^\prime}^2 -2\kappa\xi B B^\prime A^\prime-7\kappa\xi B^2 A^{\prime\prime} -\kappa\xi B B^{\prime\prime}=0,
	\end{aligned}
\end{equation}
\end{subequations}
where the prime stands for derivative with respect to $y$.
Apart from that, the bumblebee and scalar field equations \eqref{eqmat} and \eqref{eomphig} become
\begin{align}
   \label{einsteinb}
  &\lambda_B B^3-\lambda_B b^2 B+ 4\xi BA^{\prime\prime}+4\xi B{A^\prime}^2 +2\beta B {\phi^{\prime}}^2 = 0,\\
  &\phi^{\prime\prime}+4A^\prime\phi^\prime-U_\phi -4\beta B B^{\prime}\phi^{\prime}-2\beta B^2 \phi^{\prime\prime}-8\beta B^2 \phi^{\prime}A^{\prime}=0.
\end{align}
We follow the lines of Ref.~\cite{dewolfe} and take the scalar field to connect two minima of the potential $U(\phi)$ asymptotically. There, since there was no bumblebee field involved, the minimum value of the potential $U(\phi)$ could be interpreted as a cosmological constant. Here, however, this cannot be ensured anymore, as the bumblebee field may also contribute, i.e., $\Lambda_5 = U(\phi) - V(B^{\mu}B_{\mu}\pm b^2)$ for $y\to\pm\infty$.

Notice that the case $\xi=\beta=0$ and $B^2 = b^2$ leads to the standard braneworld scenario \cite{dewolfe}, described by $A^{\prime\prime} = -\kappa {\phi^{\prime}}^2/3$ and $U(\phi) = {\phi^{\prime}}^2/2 - 6 {A^{\prime}}^2/\kappa$,
which can be combined to obtain the field equation
\begin{equation}
\label{eomstd}
\phi^{\prime\prime}+4A^\prime\phi^\prime-U_\phi =0.
\end{equation}

Turning our attention to the general case, we must take into account Eq.~\eqref{einsteinb}. It admits two solutions. The simplest one is $B(y)=0$, which is a local maximum point that leads to $V(B^{\mu}B_{\mu}\pm b^2)=\lambda_B b^4/4$. Actually this is not a real vacuum, however, we present it here to illustrate the first-order formalism in a simpler situation. By considering it in Eqs.~\eqref{einsteineq}, we get
\begin{subequations}
	\begin{align} \label{all1}
  A^{\prime\prime}   &= -\frac{\kappa}3 \phi^{\prime2}, \\ \label{upa1}
  U(\phi) &=\frac12 \phi^{\prime2} - \frac{6}\kappa A^{\prime2} + \frac{\lambda_B b^4}{4}.
\end{align}
\end{subequations}
In the above expressions, the parameter $\beta$ is absent, as expected, since its corresponding term in the action \eqref{sm} vanishes. Furthermore, the field equation \eqref{eomstd} also holds in this situation. The term $\lambda_B b^4/4$ in $U(\phi)$ cancels out in the action with the potential $V$ associated with the bumblebee field. 

\section{First order formalism}\label{sec3}

To develop a first order framework, we introduce an auxiliary function, $W=W(\phi)$, and take \cite{dewolfe,cvetic}
\begin{equation}\label{phiprime}
    \phi^\prime = \frac12 W_\phi.
\end{equation}
By substituting it in Eq.~\eqref{all1}, we get that the warp function is described by
\begin{equation}\label{aprime1}
	A^\prime = -\frac{\kappa}{6}  W(\phi).
\end{equation}
In order to obtain the potential, we use Eq.~\eqref{upa1}, which leads us to
\begin{equation}\label{potw1}
U(\phi)=\frac18 W_\phi^2 - \frac{\kappa}{6}  W^2  + \frac{\lambda_B b^4}{4}.
\end{equation}
For instance, we can take the auxiliary function as
\begin{equation}\label{wphi4}
  W(\phi)=2\sqrt{\lambda_\phi}  \left(\phi_0^2{\phi}-\frac{\phi^3}{3}\right),
    \end{equation}
where $\lambda_\phi$ and $\phi_0$ are positive parameters, such that the potential in Eq.~\eqref{potw1} reads
\begin{equation}\label{potb0}
    U(\phi)=\frac12 \lambda_\phi \left(\phi^2_0-\phi^2\right)^2 - \frac{2\lambda_\phi \kappa}{3}\left(\phi_0^2{\phi}-\frac{\phi^3}{3}\right)^2+ \frac{\lambda_B b^4}{4}.
\end{equation}
Its minima are located at $\phi=\pm\phi_0$, such that $U(\pm\phi_0)=- 8\lambda_\phi \kappa/\phi_0^6/27 + \lambda_B b^4/4$. The cosmological constant, then, is $\Lambda_5 = - 8\lambda_\phi \kappa/\phi_0^6/27$.
    
The choice in Eq.~\eqref{wphi4} makes Eq.~\eqref{phiprime} be written as
\begin{equation}\label{fophi}
  \phi^\prime = \sqrt{\lambda_\phi}  \left(\phi^2_0-\phi^2\right).
\end{equation}
This equation is well-known in the literature, as it describes the so-called $\phi^4$ model in $(1,1)$ flat spacetime dimensions. The solution that connects the minima of the potential is
    \begin{equation}\label{solphi4}
    \phi(y)=\phi_0 \tanh\left({\sqrt{\lambda_\phi}\,}\phi_0\,y \right).    
    \end{equation}
    Notice that the height of the solution is controlled by $\phi_0$ and the width is driven by both $\phi_0$ and $\lambda_\phi$. By substituting the above expression into Eq.~\eqref{aprime1}, we get
  \begin{equation}
  \begin{split}
  A^\prime &= -\frac{\kappa\sqrt{\lambda_\phi}}{3} \left(\phi_0^2{\phi}-\frac{\phi^3}{3}\right) \\
  	  &=-\frac{\kappa\phi_0^3\sqrt{\lambda_\phi}}{3}\! \left(\!\tanh\!\left(\!\sqrt{\lambda_\phi}\phi_0\,y \!\right)\!   -\!\frac{1}{3}\!\tanh^3\!\left(\!{\sqrt{\lambda_\phi}}\phi_0\,y \!\right)\!   \right).
  \end{split}
\end{equation}  
This equation admits the analytical solution 
  \begin{equation}\label{aphi4}
         A(y) \!=\! -\frac{\kappa\phi_0^2}{18}\! \left({\tanh^2({\sqrt{\lambda_\phi}}\phi_0\,y )}-4\ln(\sech({\sqrt{\lambda_\phi}}\phi_0\,y ) )  \right), 
\end{equation}
where we have assumed $A(0)=0$ to determine the constant of integration. The warp factor associated with the above function goes to zero asymptotically and has a maximum at $y=0$. Notice that the parameter $b$, the VEV of the bumblebee field, is absent in Eqs.~\eqref{solphi4} and \eqref{aphi4}. It only appears in the potential \eqref{potb0}, vanishing in the cosmological constant, as this term cancels out with the contribution of the bumblebee potential.

In Eqs.~\eqref{all1}--\eqref{aphi4}, we have investigated the case $B(y)=0$. We now study a richer possibility that arises from Eq.~\eqref{einsteinb}. It is given by
\begin{equation}\label{bumbleb}
B^2(y)=b^2 - \frac{4\xi}{\lambda_B} 
\left(A^{\prime2}+A^{\prime\prime}\right) - \frac{2\beta}{\lambda_B}{\phi^\prime}^2.
\end{equation}
The above equation constrains $B(y)$ in terms of the warp function and the scalar field. To get a pair of equations depending only on $A(y)$ and $\phi(y)$, one must substitute the above expression in Eqs.~\eqref{einsteineq}. As it follows from various experimental measurements, the LV constant vectors (tensors) are known to be very small in four dimensions \cite{datatables}, and therefore it is natural to assume them to be small in higher dimensions as well, since it is supposed that the Lorentz symmetry is a fundamental symmetry of nature, even in theories defined in higher-dimensional spacetimes \cite{Kostelecky:1988zi}, see also \cite{CarrollTam} and references therein.

Therefore, it suffices to keep only the first order in these constants. Thus, throwing away higher-order terms in $\xi$, $\beta$ and their combinations, one gets
\begin{subequations}
	\begin{align} \label{all2}
  A^{\prime\prime}   &= -\frac{\kappa(1+b^2(\kappa\xi-2\beta) )}3 \phi^{\prime2}, \\  \label{vpa2}
  U(\phi) &=\frac{1-8b^2\beta}{2}\, \phi^{\prime2} - \frac{6\,(1-\kappa b^2\xi)}\kappa A^{\prime2}.
\end{align}
\end{subequations}
Notice that the parameter $\lambda_B$ does not appear in the above equations. This only occurs because we are taking into account the leading terms in the coupling constants $\xi$ and $\beta$. Furthermore, in this situation, the bumblebee potential vanishes, i.e., $V(B^\mu B_\mu \pm b^2)=0$, despite the function $B(y)$ associated with the bumblebee field is not in the minimum of $V$.

For brane solutions connecting two $AdS$ geometries, one has $\phi^\prime(\pm\infty)\to0$ and $A^\prime(\pm\infty)= A^\prime_\pm$, with $A_\pm$ being negative constants. Thus, from Eq.~\eqref{vpa2}, it is straightforward to see that the potential $U(\phi)$ goes to constant values asymptotically, which makes the cosmological constant being written as $\Lambda_5  = -6\,(1-\kappa b^2\xi) {A^\prime_\pm}^2/\kappa$. Furthermore, the conditions under $A(y)$ and $\phi(y)$ allow us to conclude from Eq.~\eqref{bumbleb} that, for $y\to\pm\infty$, the behavior of $B^2(y)$ is
\begin{equation}\label{basy}
	B^2_\pm \equiv \lim_{y\to\pm\infty} B^2(y) = b^2-\frac{4\xi A^\prime_\pm}{\lambda_B}.
\end{equation}
This shows that, for $\xi\neq0$, $B^2(y)$ does not connect the VEVs of $V$.

\begin{figure}[t!]
\includegraphics[width=0.45\linewidth]{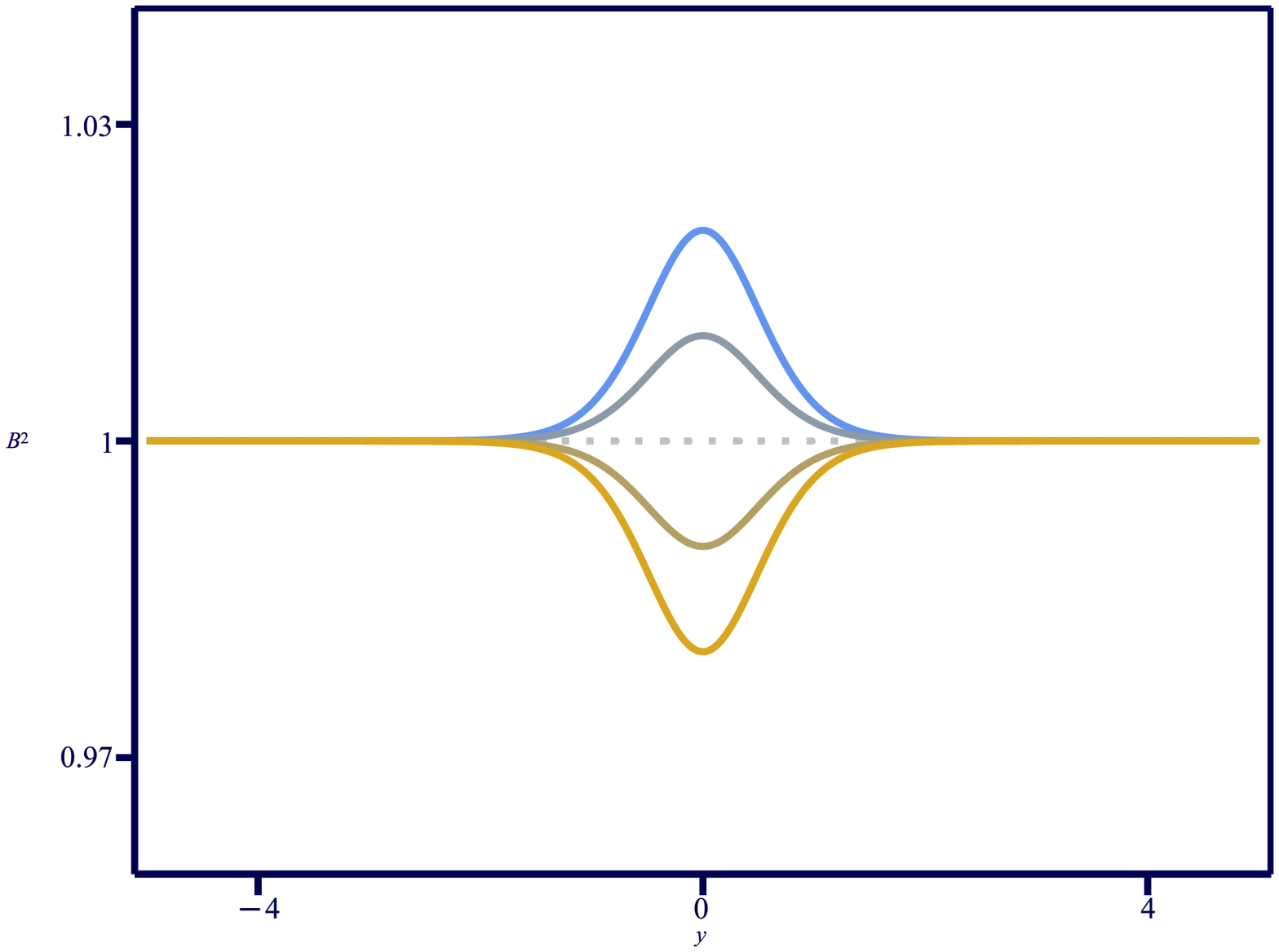}
\includegraphics[width=0.45\linewidth]{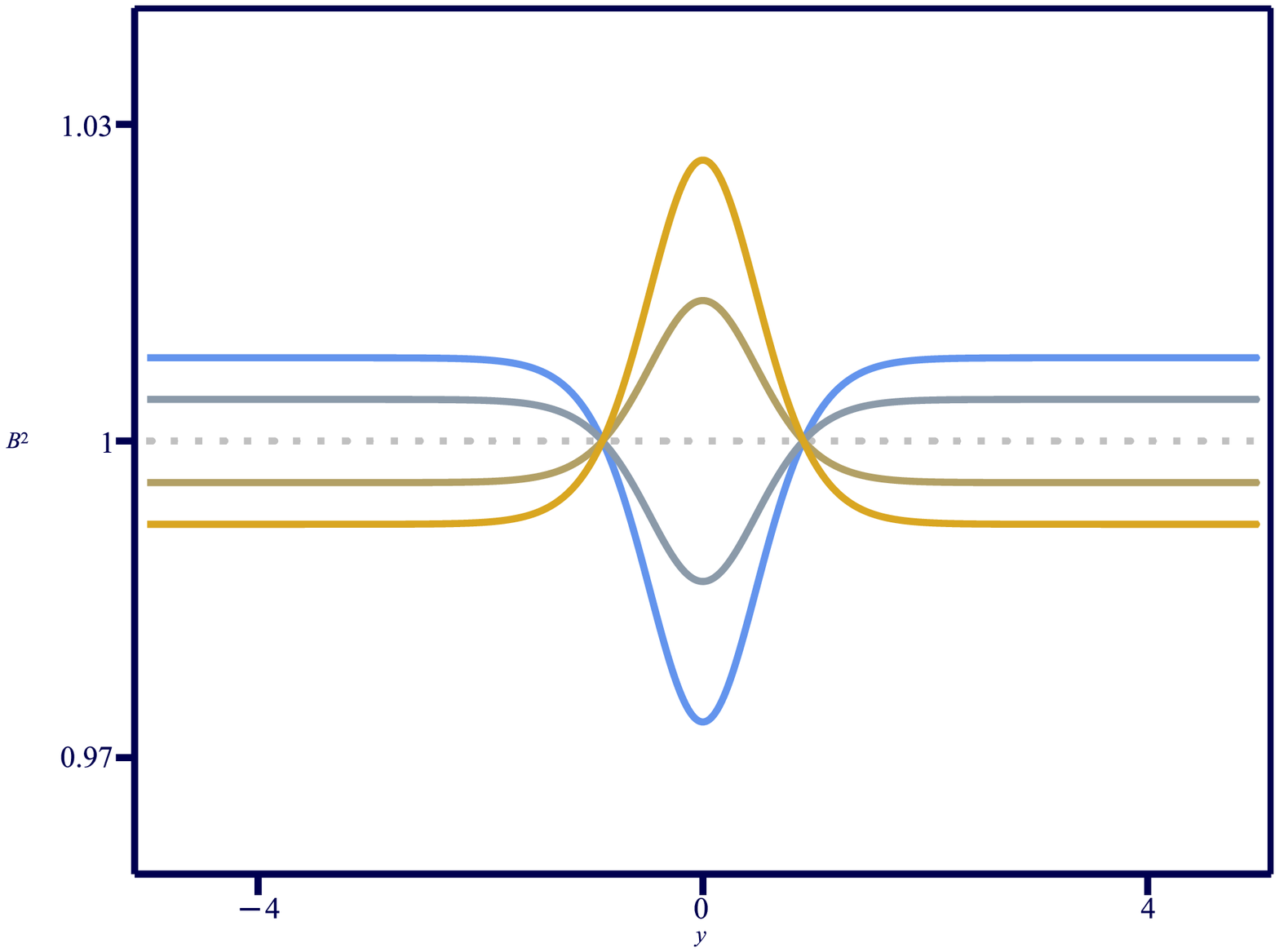}
\centering
	\caption{The solution $B^2(y)$ with $\kappa=2$, $b^2=\phi_0=\lambda_\phi=\lambda_B=1$, for $\xi=0$ and $\beta=-0.01,-0.005,0.005$ and $0.01$ (left), $\xi=-0.01,-0.005,0.005$ and $0.01$ and $\beta=0$ (right). The color of the lines ranges from blue to yellow with the increasing of the parameter that changes in each panel.}
	\label{fig1}
\end{figure}

With the aim of developing a first order framework, we use the same approach considered previously, with the auxiliary function $W=W(\phi)$ introduced as in Eq.~\eqref{phiprime}. From Eq.~\eqref{all2}, we get that the warp function becomes
\begin{equation}\label{aprime2}
	A^\prime = -\frac{\kappa}{6} (1+b^2(\kappa\xi-2\beta)) W(\phi)
\end{equation}
and the potential in Eq.~\eqref{vpa2} reads
\begin{equation}\label{uw2}
U(\phi)=\frac{1-8b^2\beta}{8} W_\phi^2 - \frac{\kappa(1+b^2(\kappa\xi-4\beta))}{6}  W^2.
\end{equation}
In order to illustrate our procedure, we take the $W(\phi)$ in Eq.~\eqref{wphi4}. In this case, the above potential takes the form
\begin{equation}
\begin{aligned}
	U(\phi)&=\frac{1-8b^2\beta}2 \lambda_\phi \left(\phi^2_0-\phi^2\right)^2\\
	&- \frac{2\lambda_\phi \kappa}{3} {(1+b^2(\kappa\xi-4\beta))}   \left(\phi_0^2{\phi}-\frac{\phi^3}{3}\right)^2.
\end{aligned}
\end{equation}
One can show that its minima are located at $\phi=\pm\phi_0$, such that
\begin{equation}
   \Lambda_5\equiv U(\pm\phi_0)=- \frac{8\lambda_\phi \kappa}{27} {(1+b^2(\kappa\xi-4\beta))}\phi_0^6.
\end{equation}
We highlight that, contrary to the previous case, $\Lambda_5$ is given only in terms of $U(\phi)$ asymptotically, since $V$ vanishes. The first order equation that governs $\phi(y)$ is the same as in Eq.~\eqref{fophi}. So, $\phi(y)$ is given by Eq.~\eqref{solphi4}, exactly as in the case with $B=0$. However, the equation \eqref{aprime2} that drives the warp function has now the form
\begin{equation}
A^\prime = -\frac{\kappa\sqrt{\lambda_\phi}}3 (1+b^2(\kappa\xi-2\beta))  \left(\phi_0^2{\phi}-\frac{\phi^3}{3}\right).
\end{equation}
By using Eq.~\eqref{solphi4}, we get
 \begin{equation}
 \begin{aligned}
 	A^\prime &= -\frac{\kappa\phi_0^3\sqrt{\lambda_\phi}}3 (1+b^2(\kappa\xi-2\beta))\\
 	&\times\left({\tanh\left({\sqrt{\lambda_\phi}}\phi_0\,y \right)}-\frac{1}{3}\tanh^3\left({\sqrt{\lambda_\phi}}\phi_0\,y \right)\right).
 \end{aligned} 
\end{equation}
Notice the presence of the parameters $b$, $\beta$ and $\xi$ in this expression, which are associated with the bumblebee field in Eq.~\eqref{bumbleb}. The above equation is solved by
\begin{equation}
 \begin{aligned}
 	A(y) &= -\frac{\kappa\phi_0^2}{18} (1+b^2(\kappa\xi-2\beta))\\
 	&\times\left({\tanh^2({\sqrt{\lambda_\phi}}\phi_0\,y )   }-4\ln(\sech({\sqrt{\lambda_\phi}}\phi_0\,y ) )  \right).
 \end{aligned} 
 \label{kkk}
\end{equation}
So, the warp function behaves asymptotically as $A(y)\approx -\kappa\phi_0^2 (1+b^2(\kappa\xi-2\beta)) |y|/27$, for $y\to\pm\infty$. This shows that the brane solution interpolates two $AdS$ geometries. In this scenario, plugging Eqs.(\ref{solphi4}) and (\ref{kkk}) into Eq.~\eqref{bumbleb}, one finds 
\begin{equation}
\begin{aligned}
	B^2(y) &=b^2 - \frac{\lambda_\phi}{\lambda_B}\frac{2\phi_0^4}{81}\Bigg(
8\kappa^2\xi\phi_0^2 \\
&+3\,(27\beta-2\kappa\xi(\kappa\phi_0^2+9))\,\sech^4({\sqrt{\lambda_\phi}}\phi_0\,y ) \\
&-2\kappa^2\xi\phi_0^2\sech^6({\sqrt{\lambda_\phi}}\phi_0\,y )\Bigg).
\end{aligned}
\end{equation}
The above expression leads us to
\begin{subequations}
\begin{align}
	& B^2_\pm = b^2-\frac{16\lambda_\phi}{81\lambda_B}\kappa^2\xi\phi_0^6,\\
	& B^2(0) = b^2+\frac{4\lambda_\phi}{3\lambda_B}\left(\kappa\xi - \frac32\beta\right)\phi_0^4,
\end{align}
\end{subequations}
where we have used Eq.~\eqref{basy} to get the asymptotic behavior of $B^2(y)$. From this, we see that $B$ only goes to the VEV asymptotically in the case $\xi=0$ and $\beta\neq0$. In this situation, the parameter $\beta$ controls its behavior at the origin, so it goes from the VEV of $V$ at $y\to-\infty$, passes another point at $y=0$ and then returns to the VEV at $y\to\infty$. 
We depict this behavior for some values of $\beta$ in the left panel of Fig.~\ref{fig1}. Notice that $y=0$ is a maximum for $\beta<0$ or a minimum for $\beta>0$.
Another possibility is to consider $\beta=0$ and $\xi\neq0$. In this case, the bumblebee field does not connect the VEVs due to the contribution of $\xi$ which appears with a negative sign in $B^2_\pm$. This behavior is shown in the right panel of Fig.~\ref{fig1}. Notice that the solution always crosses the VEV at symmetric points around the origin, attaining a minimum (maximum) at $y=0$ for $\xi<0$ ($\xi>0$).
\begin{figure}[t!]
\includegraphics[width=0.45\linewidth]{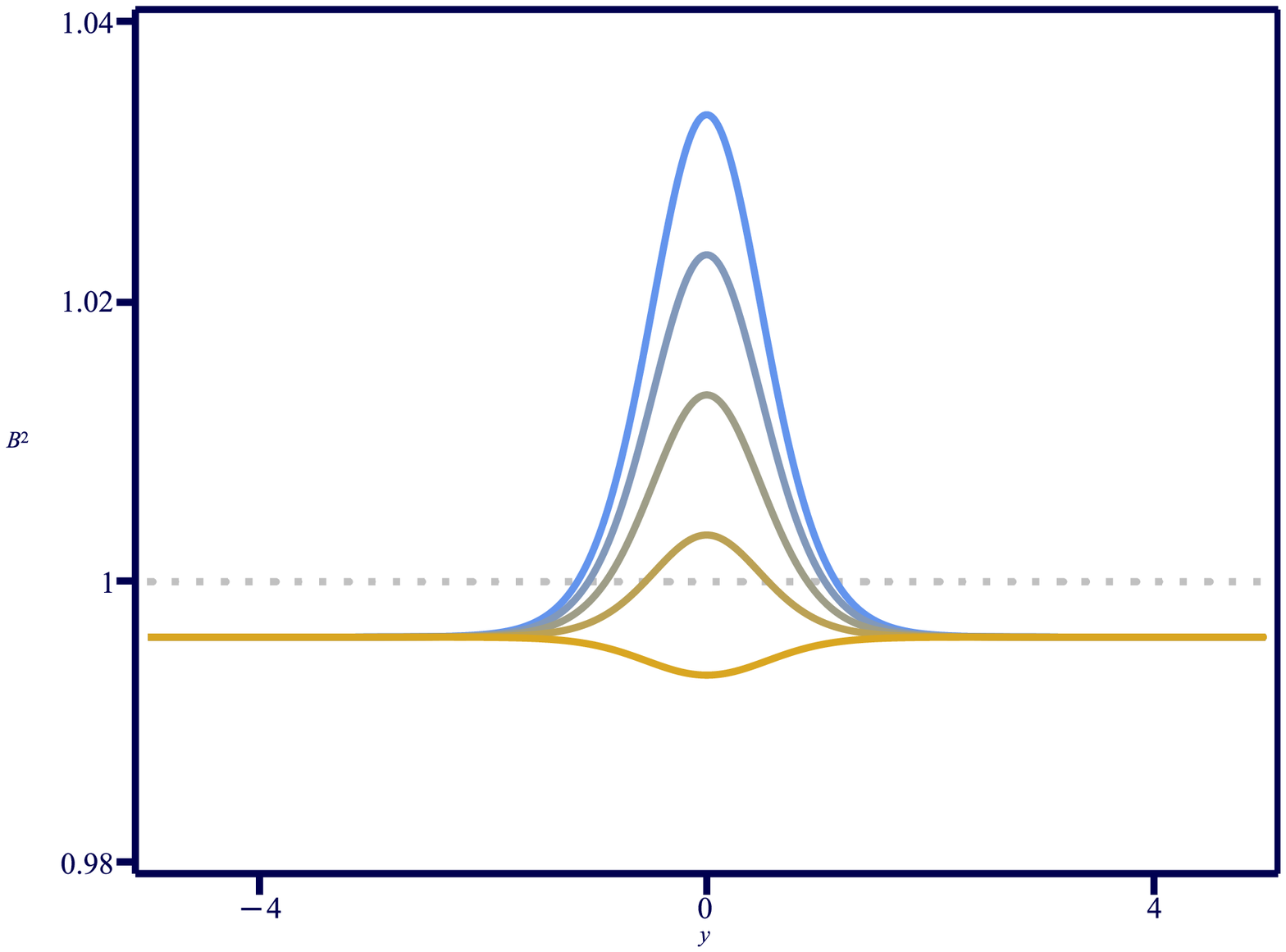}
\includegraphics[width=0.45\linewidth]{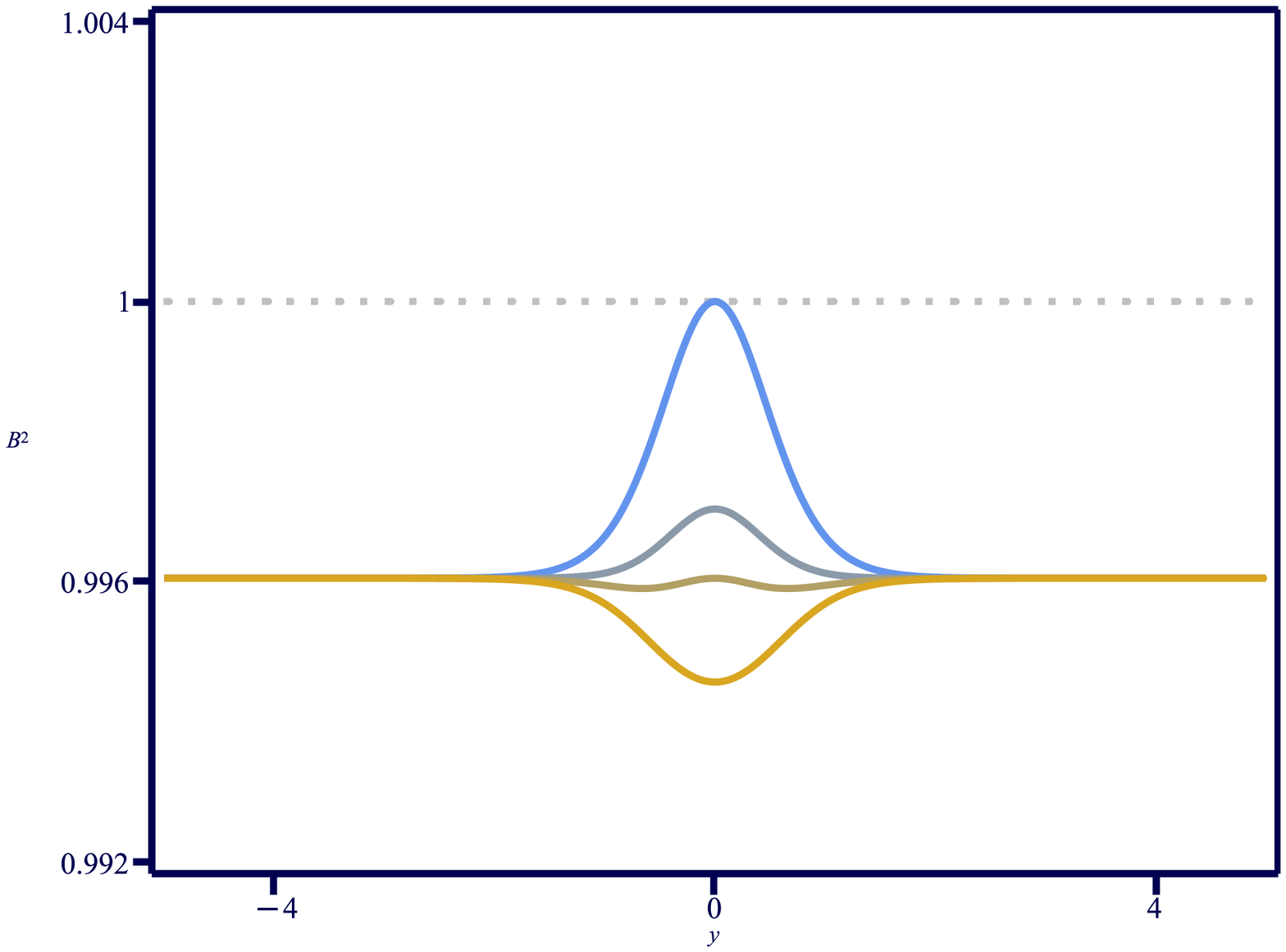}
\centering
	\caption{The solution $B^2(y)$ with $\kappa=2$, $b^2=\phi_0=\lambda_\phi=\lambda_B=1$ and $\xi=0.005$. In the left panel, we display it for $\beta= -0.01,-0.005,0,0.005$ and $0.01$. In the right panel, we plot it for $\beta = \beta_1,\beta_2,\beta_3$ and $\beta_4$. The color of the lines ranges from blue to yellow with the increasing of $\beta$.}
	\label{fig2}
\end{figure}
The case in which both parameters are not zero is richer. In Fig.~\ref{fig2}, we take $\xi=0.005$ and display it for some values of $\beta$. In the left panel, we show the general behavior, from which one can see that $\beta$ modifies the core of the solution, leading the transition from a maximum to a minimum and, also, changing the number of points that crosses the VEV. In the right panel, we highlight these features. For $\beta=\beta_1\equiv 2\kappa\xi/3$, the solution engenders a maximum that touches the VEV. As $\beta$ increases, this maximum goes down. Another value of interest is $\beta=\beta_2\equiv 2\kappa\xi(\kappa\phi_0^2+9)/27$, in which two minima arise at infinity and approaches $y=0$ as $\beta$ gets larger. In this case, $B^2(y)$ presents a maximum at its center and two minima around it. A particular situation occurs for $\beta = \beta_3\equiv2\kappa\xi(4\kappa\phi_0^2+27)/81$, in which the maximum attains the same value of the asymptotic limits, $B^2_\pm$. We continue increasing $\beta$ until the value $\beta=\beta_4 \equiv \kappa\xi(11\kappa\phi_0^2 + 54)/81$. For $\beta\geq\beta_4$ the maximum becomes a minimum. The solution is plotted for $\beta=\beta_1,\beta_2,\beta_3$ and $\beta_4$ in the right panel of Fig.~\ref{fig2}.

\vspace*{3mm}

\section{Summary}\label{sec4}
In this manuscript, we have investigated braneworlds within the five-dimensional bumblebee gravity. We have obtained the field equations for the background \ref{brane} and have shown that two possibilities arise for the bumblebee field. The simplest one is to consider that it vanishes. In this case, although the parameter $\lambda_B$ is present in the potential $U(\phi)$, the model supports a first order framework similar to the one developed in Ref.~\cite{dewolfe}. The second possibility is to consider that the bumblebee field is driven by the function in Eq.~\eqref{bumbleb}. This case is richer, with the parameters $\xi$ and $\beta$, associated with the bumblebee gravity and the presence of aether respectively, appearing in the equations. Since it is expected that the Lorentz-violating parameters are small, cf. \cite{datatables}, we have considered the case in which the aforementioned parameters are very small such that their higher-order contributions can be neglected. Under this assumption, we have shown that a first order framework can be developed. In this situation, the bumblebee field has a bell-like shape which can be deformed for specific choices for the parameters.

We hope that the current manuscript fosters other investigations concerning braneworlds in bumblebee gravity. For instance, one may consider the Lorentz symmetry-breaking scenario in which the scalar field appears with non-canonical terms in the action \cite{genbrane} or in Weyl geometries \cite{weyl}. Besides this, considering higher-curvature extensions to the pure gravity sector should be a promising way to capture the effects of the high energy regime.

\acknowledgments{This work is supported by the Brazilian agencies Conselho Nacional de Desenvolvimento Cient\'ifico e Tecnol\'ogico (CNPq), grants 306151/2022-7 (M.A.M.), 310994/2021-7 (R.M.), 301562/2019-9 (A.Yu.P) and 307628/2022-1 (P.J.P.), and Paraiba State Research Foundation (FAPESQ-PB) grants 0015/2019 (M.A.M.), 150891/2023-7  (P.J.P.) and 0003/2019 (R.M.).


\begin{thebibliography}{99}

\bibitem{KosGra} V.~A.~Kostelecky,
Phys. Rev. D \textbf{69} (2004), 105009
[arXiv:hep-th/0312310 [hep-th]].

\bibitem{Colladay1}
D.~Colladay and V.~A.~Kostelecky,
Phys. Rev. D \textbf{55}, 6760-6774 (1997)
[arXiv:hep-ph/9703464 [hep-ph]].

\bibitem{Colladay2}
D.~Colladay and V.~A.~Kostelecky,
Phys. Rev. D \textbf{58}, 116002 (1998)
[arXiv:hep-ph/9809521 [hep-ph]].



\bibitem{Shapiro} G.~de Berredo-Peixoto and I.~L.~Shapiro,
Phys. Lett. B \textbf{642} (2006), 153-159
[arXiv:hep-th/0607109 [hep-th]].

\bibitem{KosLiGrav} V.~A.~Kosteleck\'y and Z.~Li,
Phys. Rev. D \textbf{103} (2021) no.2, 024059
[arXiv:2008.12206 [gr-qc]].

\bibitem{Bert} O.~Bertolami and J.~Paramos,
Phys. Rev. D \textbf{72} (2005), 044001
[arXiv:hep-th/0504215 [hep-th]].

\bibitem{Poulis:2021nqh}
F.~P.~Poulis and M.~A.~C.~Soares,
Eur. Phys. J. C \textbf{82}, no.7, 613 (2022)
[arXiv:2112.04040 [gr-qc]].

\bibitem{Para} D.~Capelo and J.~P\'aramos,
Phys. Rev. D \textbf{91} (2015) no.10, 104007
[arXiv:1501.07685 [gr-qc]].

\bibitem{Casana} R.~Casana, A.~Cavalcante, F.~P.~Poulis and E.~B.~Santos,
Phys. Rev. D \textbf{97} (2018) no.10, 104001
[arXiv:1711.02273 [gr-qc]].

\bibitem{ourPP}
A.~A.~A.~Filho, J.~R.~Nascimento, A.~Y.~Petrov and P.~J.~Porf\'\i{}rio,
[arXiv:2211.11821 [gr-qc]].

\bibitem{godelbumb} A.~F.~Santos, A.~Y.~Petrov, W.~D.~R.~Jesus and J.~R.~Nascimento,
Mod. Phys. Lett. A \textbf{30} (2015) no.02, 1550011
[arXiv:1407.5985 [hep-th]].

\bibitem{godelbumb1} W.~D.~R.~Jesus and A.~F.~Santos,
Int. J. Mod. Phys. A \textbf{35} (2020) no.09, 2050050
[arXiv:2003.13364 [gr-qc]].


\bibitem{RS1} L.~Randall and R.~Sundrum, 
Phys. Rev. Lett. \textbf{83} (1999), 3370
[arXiv:hep-ph/9905221 [hep-ph]].

\bibitem{RS2} L.~Randall and R.~Sundrum,
Phys. Rev. Lett. \textbf{83} (1999), 4690
[arXiv:hep-th/9906064 [hep-th]].
\bibitem{garriga}J.~Garriga and T.~Tanaka, 
Phys. Rev. Lett. \textbf{84} (2000), 2778
[arXiv:hep-th/9911055 [hep-th]].

\bibitem{cvetic} M.~Cvetic, S.~Griffies and S.~J.~Rey, 
Nucl. Phys. B \textbf{381} (1992), 301
[arXiv:hep-th/9201007 [hep-th]].
\bibitem{dewolfe}O.~DeWolfe, D.~Z.~Freedman, S.~S.~Gubser and A.~Karch,
Phys. Rev. D \textbf{62} (2000), 046008
[arXiv:hep-th/9909134 [hep-th]].
\bibitem{csaki}C.~Csaki, J.~Erlich, T.~J.~Hollowood and Y.~Shirman, 
Nucl. Phys. B \textbf{581} (2000), 309
[arXiv:hep-th/0001033 [hep-th]].
\bibitem{gremm}M.~Gremm, 
Phys. Lett. B \textbf{478} (2000), 434
[arXiv:hep-th/9912060 [hep-th]].
\bibitem{brito}F.~Brito, M.~Cvetic and S.~Yoon, 
Phys. Rev. D \textbf{64} (2001), 064021
[arXiv:hep-ph/0105010 [hep-ph]].
\bibitem{kobayashi}S.~Kobayashi, K.~Koyama and J.~Soda, 
Phys. Rev. D \textbf{65} (2002), 064014 
[arXiv:hep-th/0107025 [hep-th]].
\bibitem{dzhunushaliev}
V.~Dzhunushaliev, V.~Folomeev and M.~Minamitsuji,
Rept. Prog. Phys. \textbf{73} (2010), 066901
[arXiv:0904.1775 [gr-qc]].

\bibitem{b1}D.~Bazeia and A.~R.~Gomes, 
JHEP \textbf{05} (2004), 012
[arXiv:hep-th/0403141 [hep-th]].
\bibitem{b2}D.~Bazeia, C.~Furtado and A.~R.~Gomes,
JCAP \textbf{02} (2004) 002
[arXiv:hep-th/0308034 [hep-th]].
\bibitem{b3}
D.~Bazeia, F.~A.~Brito and L.~Losano,
JHEP \textbf{11} (2006) 064
[arXiv:hep-th/0610233 [hep-th]].
\bibitem{b4}D.~Bazeia, A.~S.~Lobao, L.~Losano and R.~Menezes, 
Phys. Rev. D \textbf{88} (2013), 045001 [arXiv:1306.2618 [hep-th]]
\bibitem{b5} D.~Bazeia, R.~Menezes and R.~da Rocha, 
Adv. High Energy Phys. \textbf{2014} (2014), 276729 
[arXiv:1312.3864 [hep-th]].
\bibitem{b6}
D.~Bazeia, L.~Losano, M.~A.~Marques and R.~Menezes,
Phys. Lett. B \textbf{736} (2014), 515
[arXiv:1407.3478 [hep-th]].
\bibitem{b7}D.~Bazeia, M.~A.~Marques and R.~Menezes, 
Phys. Rev. D \textbf{92} (2015), 084058
[arXiv:1510.04578 [hep-th]].
\bibitem{b8}D.~F.~S.~Veras and C.~A.~S.~Almeida,
Phys. Rev. D \textbf{95} (2017), 104032
[arXiv:1702.06263 [gr-qc]].
\bibitem{b9}D.~Bazeia and D.~C.~Moreira,
Eur. Phys. J. C \textbf{77}, (2017) 884
[arXiv:1703.06363 [hep-th]].

\bibitem{CarrollTam} S.~M.~Carroll and H.~Tam,
Phys. Rev. D \textbf{78} (2008), 044047
[arXiv:0802.0521 [hep-ph]].

\bibitem{aether} M.~Gomes, J.~R.~Nascimento, A.~Y.~Petrov and A.~J.~da Silva,
Phys. Rev. D \textbf{81} (2010), 045018
[arXiv:0911.3548 [hep-th]].












\bibitem{classMetAf}
A.~Delhom, J.~Nascimento, G.~J.~Olmo, A.~Y.~Petrov and P.~Porf\'{i}rio,
Eur. Phys. J. C {\bf 81} (2021), 287.
[arXiv:1911.11605 [hep-th]].

\bibitem{qMetAf1} A.~Delhom, J.~R.~Nascimento, G.~J.~Olmo, A.~Y.~Petrov and P.~J.~Porf\'\i{}rio,
Phys. Lett. B \textbf{826} (2022), 136932
[arXiv:2010.06391 [hep-th]].






\bibitem{qMetAf2} A.~Delhom, T.~Mariz, J.~R.~Nascimento, G.~J.~Olmo, A.~Y.~Petrov and P.~J.~Porf\'\i{}rio,
JCAP \textbf{07} (2022) no.07, 018
[arXiv:2202.11613 [hep-th]].


\bibitem{datatables} V.~A.~Kostelecky and N.~Russell,
Rev. Mod. Phys. \textbf{83} (2011), 11-31
[arXiv:0801.0287 [hep-ph]].

\bibitem{Kostelecky:1988zi}
V.~A.~Kostelecky and S.~Samuel,
Phys. Rev. D \textbf{39} (1989), 683.


\bibitem{genbrane}D.~Bazeia, A.~R.~Gomes, L.~Losano and R.~Menezes, 
Phys. Lett. B \textbf{671} (2009), 402
[arXiv:0808.1815 [hep-th]].

\bibitem{weyl}Y.~X.~Liu, X.~H.~Zhang, L.~D.~Zhang and Y.~S.~Duan, 
JHEP \textbf{02} (2008), 067
[arXiv:0708.0065 [hep-th]].


\end{thebibliography}
\end{document}